\documentstyle[prl,aps,preprint,psfig]{revtex}

\newcommand{\neu}{\tilde{\chi}^0}

\begin{document}
\draft
\preprint{TU-576}
\title{Pattern of Neutrino Oscillations in Supersymmetry with
Bilinear R-parity Violation}
\author{Fumihiro Takayama\footnote{e-mail: takayama@tuhep.phys.tohoku.ac.jp}
 and Masahiro Yamaguchi\footnote{e-mail: yama@tuhep.phys.tohoku.ac.jp}}
\address{Department of Physics, Tohoku University \\
         Sendai 980-8578, Japan}
\date{October 1999}

\maketitle
\begin{abstract}
  Neutrino masses and their mixing are studied in detail in the
  framework of a supersymmetric standard model with bilinear R-parity
  violation. In this scenario, the mixing matrix among the neutrinos
  is in a restrictive form. We find that only the small angle MSW
  solution is allowed for the solar neutrino problem when the results
  of the CHOOZ experiment are combined with the mass squared
  difference and mixing angle suggested by the atmospheric neutrino
  data.  Collider signals of this scenario are also discussed.
\end{abstract}

\clearpage

Neutrino masses and their mixing are an intriguing window to physics
beyond the Standard Model. Neutrino oscillations among massive
neutrinos have been strongly suggested by the atmospheric neutrino
data at Super-Kamiokande \cite{SuperK} as well as by the solar neutrino 
problem\cite{Solar}.

Here  we shall consider this issue in the framework of 
a supersymmetric (SUSY) extension of the
standard model \cite{Nilles}.  
Usually in the supersymmetric standard model, 
one imposes R-parity conservation to avoid 
too fast proton decay. However, it is well known that the stability
of the proton does not forbid all R-parity non-conservation, but allows 
R-parity violation such that either lepton number
or baryon number is broken, because the nucleon decays normally require
not only the baryon number violation but also the lepton number violation.

Actually it is widely recognized that the R-parity violation with the
lepton number non-conservation \cite{HallSuzuki,Bilinear} is a
promising approach to give neutrino masses and mixing without
introducing right-handed neutrinos \cite{Seesaw}. Here we shall
consider the case where the R-parity is broken only by bilinear
terms\cite{Bilinear}. This is particularly interesting because it is
compatible with the idea of grand unified theory (GUT) where quarks
and leptons typically belong to one representation of a GUT group.

An important feature of this mechanism  of generating neutrino masses is 
that one may be able to see the R-parity violation in future collider 
experiments\cite{Collider,Experiment}. 

In this paper, we shall investigate the nature of the neutrino mass matrix
in this scenario. We find that large angle solutions to the solar neutrino
deficit are not allowed when we combine the CHOOZ result\cite{CHOOZ} with the 
$\nu_{\tau}$-$\nu_{\mu}$ solution of the atmospheric neutrino problem. This 
singles out the small mixing MSW solution\cite{MSW} for the solar neutrino in this 
scenario. We will then show that the soft SUSY breaking masses should
be finely adjusted to simultaneously explain the atmospheric neutrino 
anomaly and the solar neutrino problem. We will also discuss implications
of this specific pattern of the R-parity violation at collider experiments. 

Let us first describe the model we are considering.  The particle contents of
the model are the same as those of the minimal supersymmetric standard model
(MSSM). R-parity breaking bilinear terms are assumed to exist in superpotential
\begin{eqnarray}
  W = \mu H_D H_U +\mu_i L_i H_U +Y^L_i L_i H_D E^c_i +Y^D_i Q_iH_D D^c_i 
 +Y^U_{ij} Q_i H_U U^c_j. \label{eq:superpotential}  
\end{eqnarray}
Here $H_D$, $H_U$ are two Higgs doublets, $L_i$ a $SU(2)_L$ doublet lepton, 
$E^c_i$ is a singlet lepton, $Q_i$  a doublet quark, $U^c_i$, $D^c_i$ a 
singlet quark of up and down type, respectively. Suffices $i, j$ stand
for generations. In this basis, the Yukawa coupling matrices for the charged
leptons and for the down-type quarks are assumed to be diagonal with respect
to the generation indices. 
The soft SUSY breaking terms in the scalar potential are taken to
be of the form
\begin{eqnarray}
  V^{\mathrm{soft}}=B H_D H_U+B_i \tilde{L}_i H_U
 +m_{H_D}^2H_DH_D^{\dag}+m_{H_U}^2H_UH_U^{\dag}
  +m_{HL_i}^2\tilde{L}_iH_D^{\dag}+m_{L_{ij}}^2
  \tilde{L}_i\tilde{L}_j^{\dag} +\cdots,
\end{eqnarray}
where we have written only bilinear terms explicitly. 
In this paper, we assume lepton-Higgs universality in the soft masses, namely
\begin{eqnarray}
B_i/B = \mu_i/\mu,~ 
m^2_{L_ij}=\delta_{ij}m^2_{L_i},~ 
m_{L_i}^2= m_{H_D}^2,~  
m_{HL_i}^2=0 \label{eq:lH-universality}
\end{eqnarray}
These conditions are achieved, e.g. in the minimal supergravity
\footnote{Note that in the gravity-mediated-supersymmetry-breaking
  scheme, the lepton-Higgs universality in the soft terms may be
  violated through interactions relevant above the Grand-Unified scale
  \cite{KMY}\cite{PolonskyPomarol}.} or in scenarios of gauge mediated
supersymmetry breaking (GMSB). Our motivation to consider this
universality is alignment of vacuum expectation values with the
lepton number violating bilinear terms\cite{alignment}.  
We shall return to this point
later.

Of course, the lepton-Higgs universality is not stable against radiative
corrections and we assume that Eq.~(\ref{eq:lH-universality}) holds at an
energy scale where these soft masses are given as the boundary conditions
of the renormalization group equations.

In this model, the R-parity violating terms are parameterized by
\begin{eqnarray}
  s_3&\equiv&\sin \theta_3=
\frac{\sqrt{\mu_1^2+\mu_2^2+\mu_3^2}}{\sqrt{\mu_1^2+\mu_2^2+\mu_3^2+\mu^2}},
\nonumber\\
  s_2&\equiv&\sin \theta_2=
      \frac{\sqrt{\mu_1^2+\mu_2^2}}{\sqrt{\mu_1^2+\mu_2^2+\mu_3^2}},
\nonumber\\
  s_1&\equiv&\sin \theta_1=\frac{\mu_1}{\sqrt{ \mu_1^2+\mu_2^2}}
\end{eqnarray}
Here, for simplicity, we have taken $\mu$ and $\mu_i$ to be real.

With (\ref{eq:lH-universality}) it is convenient to define a new basis
of the Higgs and lepton supermultiplets in which the all bilinear
mixing terms between the Higgs and leptons in the superpotential
 are absent:
\begin{eqnarray}
  \left(
    \begin{array}{@{\,}cccc@{\,}}
      H_D^{'} \\
      L_3^{'} \\
      L_2^{'} \\
      L_1^{'} \\
   \end{array}    
  \right)
=
  \left(
    \begin{array}{@{\,}cccc@{\,}}
      c_3 & s_3 & 0 & 0 \\
      -s_3 & c_3 & 0 &0   \\
      0 & 0 & 1 & 0 \\
      0 & 0 & 0 & 1\\
    \end{array}    
  \right)
  \left(
    \begin{array}{@{\,}cccc@{\,}}
      1 & 0 & 0 & 0\\
      0 & c_2 & s_2 & 0\\
      0 & -s_2 & c_2 & 0\\
      0 & 0 & 0 & 1\\
    \end{array}    
  \right)
  \left(
    \begin{array}{@{\,}cccc@{\,}}
      1 & 0 & 0 & 0 \\
      0 & 1 & 0 & 0 \\
      0 & 0 & c_1 & s_1 \\
      0 & 0 & -s_1 & c_1 \\
    \end{array}    
  \right)
  \left(
    \begin{array}{@{\,}cccc@{\,}}
      H_D\\
      L_{\tau}\\
      L_{\mu}\\
      L_e\\
    \end{array}    
  \right).
\end{eqnarray}
We can neglect the Yukawa couplings for the first and second generations.
This approximation yields a global $U(1)$ symmetry, resulting in one 
massless neutrino $L_1^{'}=c_1L_{\mu}-s_1L_e$.

The rotation to take the new basis gives rise to  R-parity violating trilinear 
couplings in the superpotential as well as the softly SUSY breaking terms 
in the
scalar potential. For example a Yukawa coupling for bottom quark 
is written
\begin{equation}
                Y_b Q_3 H_D D_3^c
               = Y_b Q_3( c_3H_D^{'}-s_3 L_3^{'}) D_3^c
\end{equation}
Similarly a Yukawa coupling $Y_{\tau} L_3 H_D \tau^c$ now
reads
\begin{eqnarray}
& &Y_{\tau}(c_2s_3H_D^{'}+c_2c_3L_3^{'}-s_2L_2^{'})
  (c_3H_D^{'}-s_3L_3^{'})B^c \nonumber\\
  &=& s_2s_3Y_{\tau}L_2^{'}L_3^{'}\tau^c+Y_{\tau}c_2L_3^{'}H_D\tau^c
  -Y_{\tau}s_2c_3L_2^{'}H_D^{'}\tau^c\nonumber\\
  &\simeq& Y_{\tau}L_{\tau}s_3(s_2(c_1L_{\mu}+s_1L_e))\tau^c+Y_{\tau}L_{\tau}H_D^{'}\tau_c 
\end{eqnarray}
Note that as far as $s_3$ is small, the old basis in
Eq.~(\ref{eq:superpotential}) approximates a mass eigen basis for
charged leptons. On the other hand, the new basis with prime gives in good
approximation a mass eigen basis for neutrinos, as we will see shortly.

As we mentioned before, radiative corrections violate the relation (3),
resulting in vacuum expectation values(VEVs) of sneutrinos.  
 An analysis shows that $\tilde \nu'_3$ and $\tilde \nu'_2$ develop
VEVs.  In fact the renormalization group effects induce terms 
of the form\cite{RGE}
\begin{equation}
\Delta B_3 L'_3H_U + \Delta m^2_{HL_3} L'_3 H '^{\dagger}_D+
\Delta B_2 L'_2H_U + \Delta m^2_{HL_2} L'_2 H '^{\dagger}_D,
\label{eq:corrections}
\end{equation}
which gives the following sneutrino VEVs
\begin{eqnarray}
 \langle \tilde{\nu}'_i \rangle
 &=&\frac{\Delta B_i\tan\beta+\Delta m^2_{HL_i}}  
  {m_{L_i}^2-\frac{1}{2}m_Z^2 \cos 2\beta} \langle H_D'\rangle
~~~(i=3,2)\nonumber\\ 
  \langle \tilde{\nu}'_1\rangle&=&0.
\end{eqnarray}
Note that $\Delta B_3$ and $\Delta m_{HL_3}^2$ are proportional to the bottom
Yukawa coupling squared, while $\Delta B_2$ and $\Delta m_{HL_2}^2$ are
proportional to the tau Yukawa coupling squared, and thus the ratio
\begin{equation}
 \tan \phi =\frac{\langle \tilde \nu'_2 \rangle}{\langle \tilde \nu'_3 \rangle}
\end{equation}
is generally small.

The non-vanishing VEV for the sneutrino induces mixing between the
neutrino and the neutralinos, giving rise to a small mass for it
due to the see-saw mechanism. Here it is useful to define a new basis
\begin{equation}
\nu''_3 = \nu'_3 \cos\phi +\nu'_2 \sin\phi ,~
\nu''_2 =\nu'_2 \cos\phi - \nu'_3 \sin\phi ,~
\nu''_1 =\nu'_1.
\end{equation}
Then $\nu''_3$ acquires a tree-level mass
\begin{eqnarray}
  m_{\rm{tree}}=\frac{1}{2}g_Z^2v_{\nu}^2 \sum_i(c_WO_{W\neu_{i}}
  -s_WO_{B\neu_{i}})^2\frac{1}{m_{\neu_{i}}}.
\end{eqnarray}
Here $m_{\neu_{i}}$ ($i=1, \cdots, 4$) represent neutralino masses and
$O_{B\neu_{i}}$ and $O_{W\neu_{i}}$ stand for the contamination in 
$\neu_{i}$ of Bino and Wino
components, respectively. Also
$v_{\nu}=\sqrt{\langle \tilde \nu'_3 \rangle^2
+\langle \tilde \nu'_2 \rangle^2}$.

The neutrinos acquire masses also at the one-loop level. 
The mass matrix induced from loop diagram contribution are diagonal for
the $\nu_i'$ basis. Relevant diagrams we should consider here are those
which include the trilinear R-parity violating couplings.
A one-loop diagram including scalar bottom quarks induces a following mass for $\nu'_3$:
\begin{equation}
   m_{\nu_3'}^{\rm{loop}}= \frac{(Y_b^2s_3)^2m_b}{16\pi^2}
   \frac{m^2_{\tilde{b}:LR}}{m^2_{\tilde{b}1}-m^2_{\tilde{b}2}}
   \log{\frac{m^2_{\tilde{b}1}}{m^2_{\tilde{b}2}}}
\end{equation}
Here $m_{\tilde{b}:LR}^2$ denotes a left-right mixing mass squared in
the sbottom sector, and $m_{\tilde{b}_1}$ and $m_{\tilde{b}_2}$ are
the eigenvalues of the scalar bottom quarks.  Similarly
a one-loop diagram
involving scalar tau leptons gives a mass for $\nu'_2$
\begin{equation}
 m_{\nu_2'}^{\rm{loop}}=\frac{(Y_{\tau}^2s_3s_2)^2m_{\tau}}{16\pi^2}
   \frac{m^2_{\tilde{\tau}:LR}}{m^2_{\tilde{\tau}1}-m^2_{\tilde{\tau}2}}
   \log{\frac{m^2_{\tilde{\tau}1}}{m^2_{\tilde{\tau}2}}}
\end{equation}
Generically the latter contribution gives a dominant contribution to
the mass of the second heaviest neutrino. 
The other neutrino remains massless in the approximation that the
Yukawa couplings for the first two generations are set to be zero.

Summarizing, the mass matrix is written in the $\nu_i''$ basis
\begin{eqnarray}
  m&=&
\left(
    \begin{array}{@{\,}ccc@{\,}}
      m_{tree}+m_{\nu_3'}^{\rm{loop}}\cos^2\phi
      +m_{\nu_2'}^{\rm{loop}}\sin^2\phi 
      & (m_{\nu_3'}^{\rm{loop}}-m_{\nu_2'}^{\rm{loop}})\sin\phi\cos\phi  
      &0  \\
      (m_{\nu_3'}^{\rm{loop}}-m_{\nu_2'}^{\rm{loop}})\sin\phi\cos\phi  
      &m_{\nu_3'}^{\rm{loop}}\sin^2\phi+m_{\nu_2'}^{\rm{loop}}\cos^2\phi   
      & 0  \\
      0 & 0 & 0\\
    \end{array}
\right) 
\end{eqnarray}
When diagonalizing this matrix,  the mixing angle for $(\nu_3',\nu_2')$plane 
is 
\begin{eqnarray}
  \tan 2\delta=\frac{2(m_{\nu_3'}^{\rm{loop}}-m_{\nu_2'}^{\rm{loop}})
    \sin\phi\cos\phi}
  {(m_{tree}+(m_{\nu_3'}^{\rm{loop}}-m_{\nu_2'}^{\rm{loop}})\cos2\phi)}
  \simeq \frac{2m_{\nu_3'}^{\rm{loop}}\sin\phi\cos\phi}{m_{tree}}
  \ll 1
\end{eqnarray}
Because the tree-level mass is generically larger than the one-loop
corrections, we  find that the $\nu''_i$ basis is in a good
approximation a mass eigenstate.

Thus the mixing matrix of the neutrino sector, the MNS matrix
\cite{MSW}, becomes
\begin{eqnarray}
  U_{i \alpha}&=&
\left(
    \begin{array}{@{\,}ccc@{\,}}
      U_{\tau 3} & U_{\tau 2} & U_{\tau 1} \\
      U_{\mu 3}  & U_{\mu 2}  & U_{\mu 1}  \\
      U_{e 3}    & U_{e 2}    & U_{e 1}    \\
    \end{array}
\right) 
\nonumber \\
 &=&  \left(
    \begin{array}{@{\,}ccc@{\,}}
      1 & 0 &0   \\
      0 & c_1 & -s_1 \\
      0 & s_1 & c_1\\
    \end{array}    
  \right)
  \left(
    \begin{array}{@{\,}ccc@{\,}}
      c_2 & -s_2 & 0\\
      s_2 & c_2 & 0\\
      0 & 0 & 1\\
    \end{array}    
  \right)
\left(
    \begin{array}{@{\,}ccc@{\,}}
      c_{\phi+\delta} & -s_{\phi+\delta} & 0\\
      s_{\phi+\delta} & c_{\phi+\delta} & 0\\
      0 & 0 & 1\\
    \end{array}    
  \right)   
  =
  \left(
    \begin{array}{@{\,}ccc@{\,}}
      c_{\theta} & -s_{\theta} & 0 \\
      -c_1s_{\theta} & c_1c_{\theta} & -s_1 \\
      s_1s_{\theta} & s_1c_{\theta} & c_1 \\
    \end{array}    
  \right)   
\end{eqnarray}
with $\theta=\theta_2+\phi+\delta$. Here
$i$ and $\alpha$ denote the weak current basis and the mass eigen
basis, respectively.   Note that the MNS matrix obtained is of a very
restrictive form with only two angles. 

Now we would like to determine the mixing angles by using experimental data.
Here we assume the hierarchical mass structure, namely $m_3 \gg m_2 \gg m_1$
so that $\Delta m_{32}^2 \simeq \Delta m_{31}^2 \gg \Delta m_{21}^2$, as
will be naturally realized in the model we are considering.

Let us first consider the atmospheric neutrino anomaly, which is explained
by an oscillation $\nu_{\mu}$-$\nu_{\tau}$. The transition probability is
\begin{eqnarray}
  P(\nu_{\mu}\rightarrow \nu_{\tau}) &\simeq&
           4 |U_{\mu3}|^2 |U_{\tau 3}|^2 
           \sin^2 \frac{\Delta m_{32}^2}{4E}L
\nonumber \\
    &=& 4 c_1^2 s_{\theta}^2 c_{\theta}^2 \sin^2 \frac{\Delta m_{32}^2}{4E}L.
\end{eqnarray}
A recent analysis of the atmospheric neutrino at SuperKamiokande
\cite{Super-recent} indicates the following ranges at 90\% CL:
\begin{eqnarray}
  \Delta m_{32}^2&=&(2-6) \times 10^{-3} \mbox{eV}^2 \label{eq:mass-atm}
\\
 4 c_1^2 s_{\theta}^2 c_{\theta}^2&>& 0.85.\label{eq:mixing-atm}
\end{eqnarray}

On the other hand, the CHOOZ experiment\cite{CHOOZ} gives a bound on the 
mixing angle
$U_{e 3}=s_1 s_{\theta}$ to be
\begin{equation}
   4 s_1^2 s_{\theta}^2 (1- s_1^2 s_{\theta}^2) <0.2 \label{eq:mixing-chooz}
\end{equation}
for the mass range suggested by the atmospheric neutrino problem.

Combining Eq. (\ref{eq:mixing-atm}) with Eq. (\ref{eq:mixing-chooz}),
one finds that the $s_1$ has to be small. This non-trivial relation
among $U_{e3}$,  $U_{\mu 3}$ and $ U_{\tau 3}$ in the MNS matrix
restricts patterns of the neutrino oscillations. 
In fact, 
for small $U_{e3}$ implied by small $s_1$, the three flavor solar 
neutrino oscillation is well approximated by the usual two flavor
 oscillation solutions, namely the small angle MSW, the large angle MSW, and
 the vacuum oscillation\cite{three-flavor}.
However, the latter two are excluded because the mixing angle involves $s_1$. 
This argument enables us to single out, in this model, the small angle MSW
solution as the only possible explanation in terms of the neutrino
oscillations to the solar neutrino problem.  This is one of the main
results of our paper.

For the small angle MSW solution, a recent result for the
solar neutrino requires \cite{Solar}
$  4 c_{\theta}^2 s_1^2 c_1^2 \simeq (0.3-1) \times 10^{-2}$,
which leads to
\begin{equation}
   s_1 \simeq 0.04 - 0.07
\end{equation}

We shall next discuss the masses of the neutrinos. In general, the
loop-level mass is suppressed compared to the tree-level mass simply
due to the loop factors, giving a huge hierarchy among the neutrino
masses. The atmospheric neutrino indicates the mass of the heaviest
neutrino in the range of Eq.~(\ref{eq:mass-atm}).  The small angle MSW
solution to the solar neutrino requires \cite{Solar}
\begin{equation}
   \Delta m^2_{\rm{SMSW}} \simeq (0.4-1)\times  10^{-5} \mbox{eV}^2.
\end{equation}
To explain the atmospheric neutrino anomaly and the
solar neutrino problem simultaneously in this framework of the 
R-parity violation, one has to make the tree-level mass relatively small,
which requires a partial cancellation in the  
$\Delta B_3\tan\beta+\Delta m^2_{HL_3}$ term. 

In the following, we shall demonstrate this fine tuning in the context
of the GMSB \cite{DineNelson,INTY}. Similar conclusions will be
obtained for other mechanisms of supersymmetry-breaking mediation,
such as the gravity mediation.  In a simple GMSB model where the
messenger quark belongs to a pair of 5 and $\bar 5$ representations,
the masses of squarks and gauginos are characterized by one parameter,
the effective SUSY breaking scale $\Lambda$. The universal trilinear
$A$ term as well as the universal $B$ term are assumed to be
generated.  All these parameters are given at some energy scale, which
we identify rather arbitrarily with $\Lambda$ $(=10^5$ GeV) in the
following analysis. We also introduce the supersymmetric $\mu$ term
and $\mu_i$ ($i=$1, 2, 3) as independent parameters. Given $\tan
\beta$, defined as the ratio of the two VEVs of the neutral Higgs bosons, 
 the conditions of the
radiative electroweak symmetry breaking fix the values of $\mu$ and
$B$. The value $s_3$ is determined so that the heaviest neutrino mass
lies in the region suggested by the atmospheric neutrino anomaly. Here
we take its representative value to be $m_{\nu_3}=0.06$ eV. As was
explained earlier the other parameters $s_2$ and $s_1$ characterize
the neutrino mixing angles.

In Figs. 1 and 2, we plot contours of the ratio of the two
neutrino masses, $R=m_{\nu_2}/m_{\nu_3}$, in the $\tan \beta$ -$A$
plane. Fig. 1  represents the case where $\mu<0$ and Fig.2  corresponds
to the case of $\mu>0$.  One finds that the preferred ratio
$R\simeq 0.03-0.07$ for the MSW solution to the solar neutrino is realized
in small regions of the parameter space. We  checked numerically
that in these regions a cancellation among the two contributions to the
VEV of the sneutrino indeed occurs.

We also examined the value of $s_3$. We found that the typical
values of the $s_3$ is large, compared to a naive expectation of
$(m_{\nu_3}/m_W)^{1/2}\sim 10^{-6}$.  Especially in the region where
the MSW solution is obtained, we checked that the value of $s_3$ can
be $\sim 10^{-3}-10^{-5}$. These observations confirm that the alignment of
the vacua works at least partially so that one does not need to assume
a priori a very small $s_3$ to obtain small neutrino masses. Of course to
simultaneously obtain the solutions of the atmospheric neutrino as
well as the solar neutrino, we need the additional fine tuning to
further suppress the sneutrino VEV.

Let us now turn to the question of how one can test this scenario at
future colliders. When the R-parity is violated, the lightest
superparticle (LSP) which is also the lightest R-parity odd particle
is no longer stable, but decays to ordinary particles. In this paper
we want to discuss patterns of the decays of the LSP. Here we assume
that the LSP is a neutralino.\footnote{In the GMSB models, actually
  the LSP is the gravitino. However, one finds that the decay of the
  neutralino into the gravitino is negligibly small, and thus one can
  concentrate on the R-parity violating decay of the {\em LSP} among
  the MSSM superparticles.} Indeed in the model described above a
bino-dominant neutralino becomes the LSP.  For simplicity we also
assume that it is heavier than the W boson. The LSP then decays to the
W (Z) boson and a charged lepton (a neutrino) through the sneutrino
VEV.  Or it decays to three body decays via scalar lepton exchanges
using the trilinear R-parity violating couplings.  In either case, we
checked that the decay length is short enough for the LSP to decay
inside a detector for typical parameters discussed earlier. In Fig. 3,
we exhibit the ratio of the partial width of the two body decay which
involves the vector boson to that of the three body decays,
$\Gamma(\tilde \chi^0_1 \rightarrow \mbox{three
  leptons})/\Gamma(\tilde \chi^0_1 \rightarrow W l_i, Z \nu_i)$.  We
find that for the MSW solution the two body decays are generally
dominant decay modes, though the decays through the trilinear Yukawa
couplings may not be negligibly small.

Detailed analysis of the decays of the LSP may provide a test of the
scenario we are discussing here. Let us illustrate this again when the
LSP is bino-like. For the two body decay to a W boson and a charged
lepton, the generation structure of the charged lepton directly
reflects the mixing angles in the neutrino sector, and thus
\begin{equation}
\frac{\Gamma (\tilde \chi^0_1 \rightarrow \mu W)}
{\Gamma (\tilde \chi^0_1 \rightarrow \tau W)}= 
\frac{|U_{3\mu}|^2}{|U_{3\tau}|^2}=\tan^2 \theta.
\end{equation}
This will give information on the mixing angle of the atmospheric neutrino 
in this R-parity violation scenario.
Note that the above relation is very generic, which is valid even when one
relaxes the lepton-Higgs universality among the soft masses.

On the other hand, the three body decays occur through
the following Yukawa couplings,
\begin{equation}
Y_{\tau} s_2 s_3 c_1 (\nu_{\tau} \mu -\nu_{\mu} \tau)\tau^c
+Y_{\tau} s_2 s_3 s_1 (\nu_{\tau} e - \nu_e \tau) \tau^c
\end{equation}
The neutralino LSP, which is in our case bino-like, couples most strongly
to the right-handed stau and tau lepton and thus the decay occurs via the
right-handed stau exchange and the final state will be either 
$\tau \tau \nu_i$ or $\tau \nu_{\tau} l_i$ where the generation suffix $i$
represents the first two generations.  The latter final state is interesting
because the tagging of the charged leptons
in the first two generations will enable us to determine one parameter $s_1$.
In fact we find
\begin{equation}
   \frac{\Gamma(\tilde \chi^0_1 \rightarrow \tau \nu e)}
{\Gamma(\tilde \chi^0_1 \rightarrow \tau \nu \mu)} = \tan^2 \theta_1.
\end{equation} 
This may provide us with an important information on our scenario where
a large $s_1$ is excluded by the combined use of the CHOOZ experiment
and the atmospheric neutrino. 

To summarize, we have discussed the case of bilinear R-parity violation
as a source of neutrino masses and mixing. We found that the resulting
mixing matrix of the neutrinos has a very special pattern. This leads us to 
conclude that the large mixing angle solutions to
the solar neutrino problem are ruled
out when the CHOOZ result and the atmospheric neutrino data are combined
together. Furthermore the relatively less hierarchical structure of the
neutrino masses in this case are obtained if the soft SUSY breaking masses
are suitably tuned to give small VEV for sneutrinos. Finally we argued that
the analysis of the decays of the neutralino may reveal information on the
mixing angles among the neutrinos. 
This scenario is testable at both neutrino oscillation experiments,
{\it e.g.} SuperKamiokande\cite{SuperK2}, SNO\cite{SNO}, and KamLAND\cite{Kamland}, 
 and collider experiments in future.

\acknowledgments
This work was supported in part by the Grant-in-Aid for Scientific 
Research from the Ministry of Education, Science, Sports, 
and Culture of Japan, 
on Priority Area 707 "Supersymmetry and Unified Theory of Elementary
Particles", and by the Grant-in-Aid No.11640246 and  No.98270.

{}

\begin{figure}[t]
  \begin{center}
    ~\hfill
    \makebox[0pt]{
      \mbox{
        {
          \psfig{file=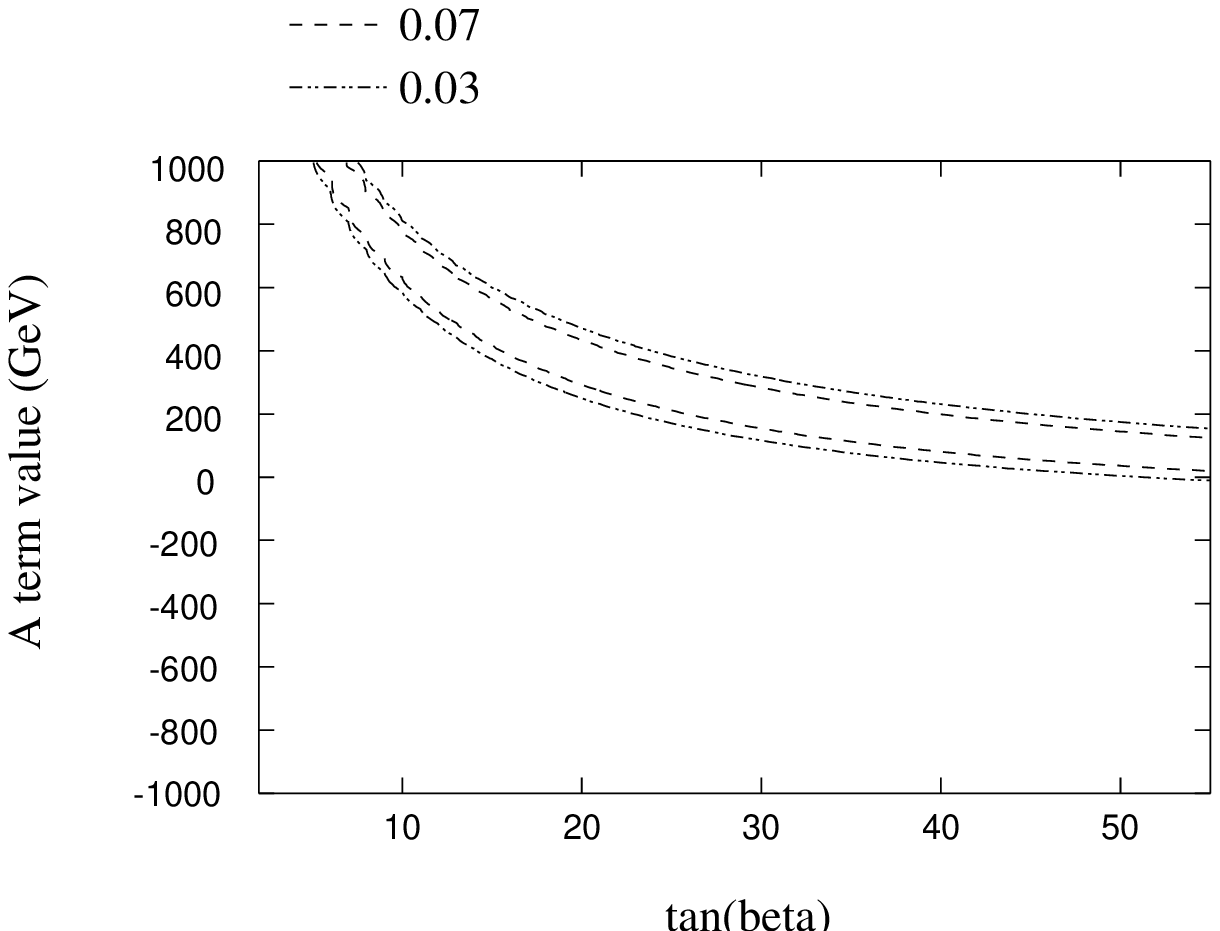}
          }
        }
      }
    \hfill~
  \end{center}
  \caption{Contour plot of $R=m_{\nu_{2}}/m_{\nu_{3}}$ on the trilinear
coupling $A$(GeV)-$\tan\beta$ plane for $\mu<0$. The dashed line corresponds 
to $R=0.07$ while the dot-dashed line corresponds to $R=0.03$.   }
  \label{charpro}
  \begin{center}
    ~\hfill
    \makebox[0pt]{
      \mbox{
        {
          \psfig{file=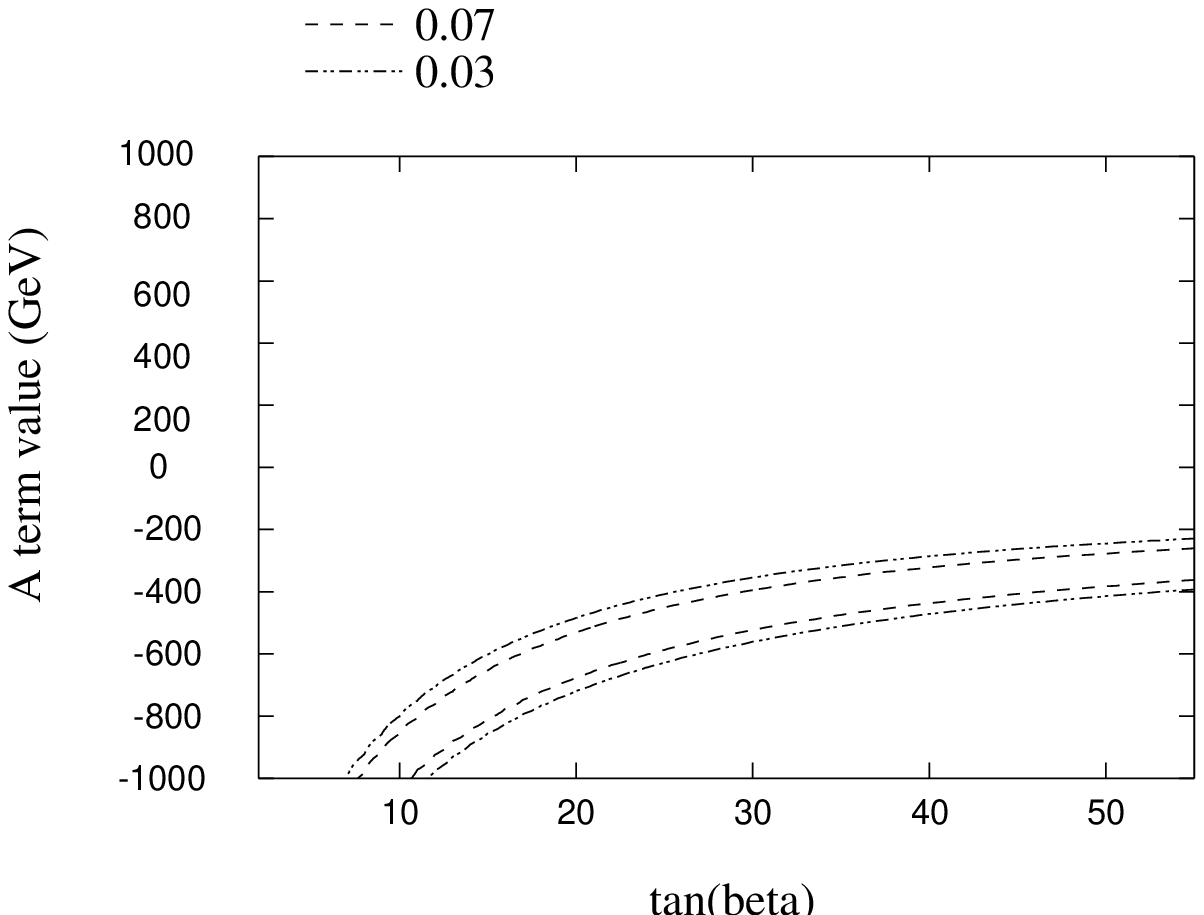}
          }
        }
      }
    \hfill~
  \end{center}
  \caption{The same figure as Fig. 1 except that $\mu>0$.}
  \label{charpro2}
\end{figure}
\begin{figure}[t]
  \begin{center}
    ~\hfill
    \makebox[0pt]{
      \mbox{
        {
          \psfig{file=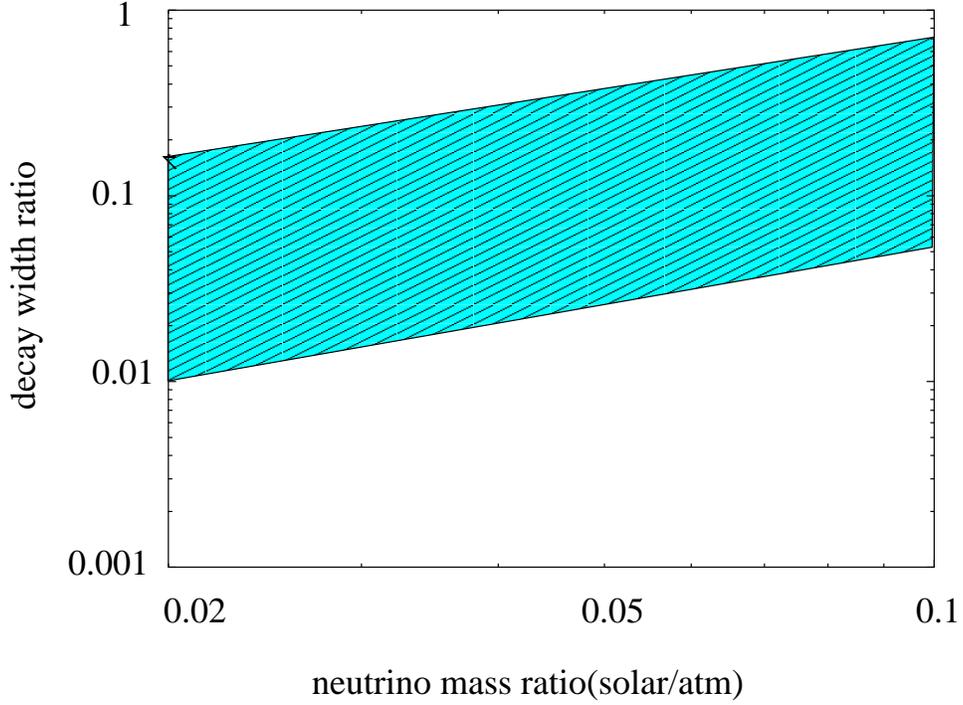}
          }
        }
      }
    \hfill~
  \end{center}
  \caption{The ratio of the decay widths $\Gamma(\tilde \chi^0_1 \rightarrow 
    \mbox{three leptons})/\Gamma(\tilde \chi^0_1 \rightarrow W l_i, Z
    \nu_i)$ versus the ratio of the neutrino
    masses $m_{\nu_{2}}/m_{\nu_{3}}$. The shaded region is obtained when $R=0.03-0.07$
is imposed.}
  \label{charpro3}
\end{figure}

\end{document}